\documentclass{sig-alternate}

\begin{document}

\setcopyright{acmlicensed}
\conferenceinfo{GECCO '15,}{July 11 - 15, 2015, Madrid, Spain}
\isbn{978-1-4503-3472-3/15/07}\acmPrice{\$15.00}
\doi{http://dx.doi.org/10.1145/2739480.2754804}

\title{Finding a Mate With No Social Skills }

\numberofauthors{2} %  in this sample file, there are a *total*

\author{
% 1st. author
\alignauthor
Chris Marriott\\
       \affaddr{University of Washington}\\
       \affaddr{1900 Commerce Street}\\
       \affaddr{Tacoma, Washington}\\
       \email{dr.chris.marriott@gmail.com}
% 2nd. author
\alignauthor
Jobran Chebib \\
       \affaddr{University of Z\"urich}\\
       \affaddr{190 Winterthurerstrasse 8057}\\
       \affaddr{Z\"urich, Switzerland}\\
       \email{jobran.chebib@ieu.uzh.ch}
}

\maketitle
\begin{abstract}
Sexual reproductive behavior has a necessary social coordination component as willing and capable partners must both be in the right place at the right time.  While there are many known social behavioral adaptations to support solutions to this problem, we explore the possibility and likelihood of solutions  that rely only on non-social mechanisms.  We find three kinds of social organization that help solve this social coordination problem (herding, assortative mating, and natal philopatry) emerge in populations of simulated agents with no social mechanisms available to support these organizations.  We conclude that the non-social origins of these social organizations around sexual reproduction may provide the environment for the development of social solutions to the same and different problems.
\end{abstract}

%A category including the fourth, optional field follows...

%\terms{Experimentation, Theory}


\section{Introduction}

Finding a mate means coordinating the location in time and space of two willing participants.  This social coordination problem (called the \emph{encounter problem} by Gimelfarb \cite{G88}) can be solved by social species like humans through social behaviors like negotiation.  There are many social behaviors that might support solutions to the encounter problem in humans and other species including herding \cite{B92, I01, R87, SB00}, philopatry \cite{CL12, G80, SG04}, pair bonding \cite{CK02}, and assortative mating \cite{ARKS96, DPSG98, G88, JBK13, K87, M91, WWVK05}.  However, sexual reproduction is a ubiquitous phenomena and a compelling hypothesis is that not all sexually reproductive life  are capable of coordinative social behaviors.  We present evidence that social coordination can emerge around sexual reproduction in populations of agents with no social behaviors.

For the purposes of our discussion we draw distinctions between social and non-social problems and social and non-social solutions to problems.  We call a problem \emph{social} if it involves the coordination of behavior of more than one individual.  Sexual reproduction is a clear social problem since agents must coordinate their behavior in some way.  This is in contrast to non-social problems that might be solved by a solitary agent without the coordination with others.  Foraging for food is an example of a non-social problem.

A solution to a problem is \emph{social} if the mechanisms that lead to the solution involve interaction between the individuals.  Specifically, the focus is on the mechanisms used in the solution to the problem and whether those mechanisms involve interaction (exchange of information) between agents.  A problem may be non-social but still have social solutions.  Foraging can be solved by a solitary agent yet in many species social solutions are observed (e.g. eusocial insects or pack hunting animals).

Coordinated behavior between agents can arise due to the social forces of social influence or social learning \cite{W00, WH92, WMMH08}.  However, it is well known that coordinated behavior can arise from non-social mechanisms as well \cite{WH92}.  This means that non-social solutions might be found for social coordination problems.  In other words the social coordination problem of sexual reproduction can potentially be solved in non-social species by non-social mechanisms.  We employed computer simulation to explore the nature of these non-social solutions following the hypothesis that sexual reproduction might serve as an evolutionary catalyst for the emergence of social behaviors in populations.

Our simulation involves agents that have no means of social interaction of any kind except sexual reproduction.  To be clear, this means that agents cannot sense the presence or actions of other agents other than when seeking a mate.  When seeking a mate they are aware only of the presence or lack of another willing mate.  Thus the agents are not capable of social interaction ruling out social solutions to the coordination problem.

Even with these limitations we show that the agents solve the social problem of sexual reproduction through social organization like herding, natal philopatry, and assortative mating.  This social coordination emerges due to the constraints that the coordination problem places on the behavior of sexual agents.

We conclude with a discussion of the role sexual reproduction plays in the emergence of social behaviors.  We propose a common pathway in which sexual agents could become socially aware.  The sexual coordination problem leads first to non-social solutions that have as their consequence social structure in the population.  This social structure can serve as the environment in which social behaviors may emerge and potentially shift the non-social solution to a social solution by relying on social signals.  We argue that the first steps down this pathway can occur if two conditions exist.  First, breeding pair selection must be influenced by the spatial distribution of individuals in the population.  Second, the spatial distribution of individuals in the population must be at least indirectly influenced by the genotype of the individuals.

\section{Mimetic Processes}

Following Whiten and Ham \cite{WH92} we use the term \emph{mimetic process} to mean ``...[any process] whereby some aspect of the behavior of one animal, B, comes to be like that of another, A''.  There are many known mimetic processes classified by Whiten and Ham in three broad categories.  The mimetic processes can be \emph{non-social} in origin, arise from \emph{social influence}, or arise from \emph{social learning}.

Since we have limited the channels of social communication between the agents we are most interested in the non-social mimetic processes.  Whiten and Ham identify four possible types of non-social mimetic processes: \emph{convergence}, \emph{common descent}, \emph{mimicry}, and \emph{individual learning}.

\emph{Convergence} is an evolutionary mechanism by which individuals in similar evolutionary niches face similar selective pressures and thus may be selected to have convergent behavior profiles.  In contrast, \emph{common descent} is when the individuals have a shared ancestry and have inherited similar behavior profiles from that ancestry.  \emph{Mimicry} is another evolutionary mechanism that selects for behaviors that are similar to another group in order to blend into that group either for protection (i.e. camouflage, defensive mimicry) or to attract prey or mates.  While this last mechanism implies an indirect ``copying'' behavior this is distinguished from social mechanisms in that mimicry is an evolutionary mechanism, not a mechanism of individual or social adaptation.

The last non-social mimetic process is not an evolutionary process.  It is an \emph{individual learning} process.  This mimetic process occurs when two learning individuals are exposed to the same or similar learning environments and thus learn similar behaviors.  This is another type of convergent process.  Like evolutionary convergence above this mechanism operates on the fact that there are convergent solutions that will be found by the adaptive process independent of interaction between individuals.  Learning convergence will play a minor role in the agents which are not capable of any significant ontogenetic development.

\subsection{Herding}

We use the term \emph{herding} broadly to mean the tendency for individuals to group together in \emph{herds} \cite{B92}, \emph{schools} \cite{I01}, \emph{flocks} \cite{R87}, \emph{swarms} \cite{SB00}, or the like.  Herding behavior can be seen as a similar behavioral phenotype among the individuals of the herd.  As a result herding behavior can be explained as occurring as a product of some mimetic process.  Herding behavior can also be seen as a mechanism underlying mimetic processes.  A herd will likely all eat, sleep, mate and migrate together and so they will be exposed to similar evolutionary and learning niches.  Herding behavior can thus act as a mechanism underlying mimetic processes of evolutionary convergence and/or individual learning convergences for behaviors like foraging, nesting, and mating.

Further many of the social mimetic processes may also be supported by a herding behavior.  Spending time with others is the precondition of social influence and social learning and thus herding behavior ensures this precondition is met.  The social mechanisms that could be supported by a herd include behavioral contagion, exposure, stimulus enhancement, imitation, over-imitation, and emulation (see \cite{WH92, WMMH08} for more details on these social mechanisms).

Herding behavior is well known to be supported by a set of very few social rules (\cite{B92, I01, R87, SB00} all use simple social rules to explain herding behavior).  This simple and elegant solution has often served as an explanation of the wide variety of herding behaviors seen in nature and has become useful in analysis of this behavior.  We wish to challenge an unspoken understanding that this elegant social solution is the only explanation of this organization.  We show that herds can also form in non-social groups due to other forces. 

\subsection{Philopatry}

\emph{Philopatry} is the tendency for an individual to remain or return to common locations during its lifetime.  \emph{Territorial philopatry} (also called \emph{territorial fidelity}) is the tendency to claim a territory or reuse specific navigation paths for travel. \emph{Breeding site fidelity} is when individuals repeatedly breed at the same location.  \emph{Natal philopatry} is a particular type of breeding fidelity when an individual returns to the place of its birth for breeding \cite{CL12, G80, SG04}.  

As we might expect philopatry will commonly co-occur in populations that also display a degree of herding.  It is hard to extract a causal relationship between this observance as we could in theory explain herding with reference to philopatric behaviors or vice versa.  This again exposes that these behaviors are both the product of mimetic processes and can serve as supporting mechanisms of other mimetic processes.

Instead of discussing causation, we focus on whether non-social or social mimetic processes are at work when considering the philopatry observed.  Some instances of philopatry may arise due to a social behavior, whereas others may be supported by non-social processes.  

\subsection{Assortative Mating}

\emph{Assortative mating}, commonly also known as \emph{homogamy}, refers to non-random mating patterns.  A random mating pattern asserts that all mating pairs are equally likely in a population.  In many cases mate selection tends to result in a sorting of mates for similarity (and occasionally dissimilarity) along a number of phenotypic or genotypic dimensions.  For instance assortative mating along size \cite{ARKS96}, age \cite{FP03}, appearance \cite{K87}, or perceived fitness \cite{K87, M91} in general are all phenotypic assortments observed in nature.  Genotypic assortment occurs when those that reproduce are closely related genetically and can often be a byproduct of phenotypic assortment.  Genotypic assortment is also commonly cited as a cause of sympatric speciation \cite{V01}.

Assortative mating is a shared behavior pattern among individuals and thus can be explained as a product of a mimetic process.  Standard explanations of assortment in nature fall into both the social and non-social categories.  While it is commonly admitted that assortment may occur due to non-social processes \cite{FP03, JBK13}, biological studies have uncovered several social mechanisms of mate preference that can lead to assortment (see \cite{JBK13} for a survey of these cases).  Nonetheless, Gimelfarb \cite{G88} shows that even when these mechanisms are in play it may still be the case that non-social forces explain the assortment.

Again, we can look for links between assortative mating and herding and philopatry.  We find co-occurrence of these behaviors in nature and simulation.  Common explanations of why species engage in dispersal of members is to avoid inbreeding that comes with philopatry \cite{CL12,G80}.  

As in herding and philopatry, we can see that assortative mating may support mimetic processes like the non-social evolutionary processes of convergence and common descent.  In individuals capable of social influence or social learning, assortative mating might lead to an assortative social learning structure as well \cite{M91}.

\section{Model}

\begin{figure}[!t]
\begin{center}
\includegraphics[width = 240pt]{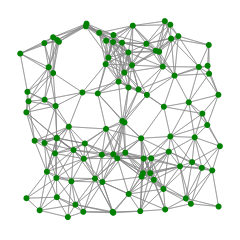}
\end{center}
\caption{A typical random geometric network.}
\label{fig:map}
\end{figure}

Our simulation consists of agents in a random geometric network \cite{P03} of resource sites.  We select random geometric networks because they are representative of real space and are useful for understanding social networks \cite{ABT14}.  See Figure~\ref{fig:map} for an example environment.  Each day the agents expend energy to move from site to site, forage for resources \cite{M00}, and engage in mating.  Energy in the simulation corresponds to the time an agent can spend doing activities.  The resources gathered determines the energy the agent has for activities in the next day.  The net daily energy gain or loss determines whether the agent lives or dies (if the energy is depleted) and whether the agent is capable of reproduction (if stored energy exceeds a threshold).

The possible behaviors of the agents are encoded in their genomes as a path of resource sites in the network (our model extends the model from \cite{MC14} and a more detailed description of the simulation is available in \cite{MC15}).  Each day the agents select a sub-path of their genome that corresponds to a path of sites that can be visited with the energy they have available for that day and that maximizes their expected resource gain.  This is carried out by locating all sites in the genome that correspond to the agent's current site and then tracing out sub-paths from the genome leading from those sites with total energy cost less than or equal to the agent's current energy.  Finally, the path that appears to maximize the resource gain for the agent is selected.  This selection depends on the current site of the agent so the behavior of the agent is determined by the coupling of its genotype with the environment.  For the purposes of our analysis this daily selected path is the relevant phenotype of each agent.

Resource gathering at each site is determined by a learning task \cite{MC14,MC15}.  Each gene in the genome of an agent corresponds to how a task is performed at a specific gathering site. Different alleles at each gene, corresponding to different ways a task is performed, determine how much energy must be used to gather resources at that site.  Each site rewards a fixed constant number of resources on each attempt (unless depleted).

The agents can engage in sexual reproduction when they have an energy total above a reproduction threshold and can find another willing and able participant at the same site.  Whether the agent will seek a mate at a site or not is also encoded at each gene in the genome.  If an agent is seeking a mate at a site it will remain at the site for an extended period after gathering resources.  If two agents are seeking mates at the same site at the same time and they are both above a sexual reproduction threshold then they will engage in sexual reproduction.  If no mate is found then agents will wait until the next opportunity to mate or eventually engage in asexual reproduction.

In this experiment both the genotype and the phenotype can be expressed as a path in the network.  We use the Levenshtein edit distance \cite{L66} to measure the distance between two paths in the graph.  This gives us a means of measuring the genotypic and phenotypic difference between agents allowing us to track mating trends and behavioral similarity.

The genotypic and phenotypic difference between mates is measured to determine the degree of assortment in the population along these dimensions.  The phenotypic difference between any two agents represents whether they visited different locations during the day.  Two agents with identical behavioral phenotypes will visit all the same sites in the same order implying a herd like structure.  We track pairwise phenotypic differences in the population and use this to track herding in the simulation.  

We measure which sites are used by an agent for breeding and use this data to measure breeding site fidelity and natal philopatry.  We measure the levels of breeding site reuse for agents and for sites over the duration of the simulation.  We also classify breeders as philopatric if they have bred at the site of their birth (at least once).  If it has only bred at sites other than its birth site then it is non-philopatric. 

\section{Observations}

We run the simulations under experimental and control conditions.  The experimental simulation consists of a population of agents all descended from a single agent.  This facilitates a close observation of the emergence of sexual reproduction in the initially small asexual population.  In contrast we compare this to control runs seeded with one hundred genetically random agents.  The initial population of control runs are related only by chance and not by descent.  This allows us to observe random mating patterns.  

\subsection{Experimental Runs}

We seed the experimental runs with a single agent created by randomly generating a genome (we generate the path by simulating a random walk on the network \cite{L93}).  This random agent is deposited in the environment and will not always be viable.  The random agent will follow the random walk encoded in its genome until it finds a local optimum that maximizes resource gain.    Assuming it can find an optimum that results in positive resource gain the agent will eventually asexually create offspring.  

With high likelihood the agent and its offspring will select the same path from their similar genomes and thus will travel together.  If along this shared path the agents spend time looking for a mate there is again a high chance of these periods overlapping leading to sexual reproduction.  If these likely events do not occur due to a disruptive mutation, then there is often another chance for this lucky event to occur soon after with the same two agents, or with future asexual offspring.  As a result the first sexual reproduction occurs soon after the first asexual offspring are born.

% herd heatmap or dendrogram or both

\begin{figure}[!t]
\begin{center}
\includegraphics[width = 240pt]{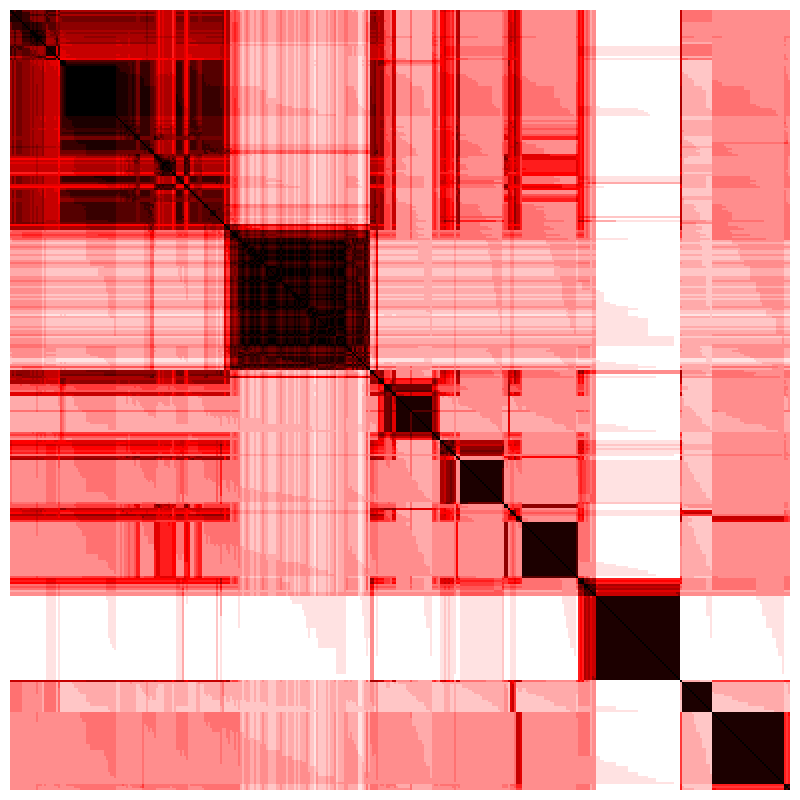}
\end{center}
\caption{Each large dark square along the diagonal represents a cluster of agents with the same daily activity, that is, a herd.  This data shows 390 agents in a population after 10000 days of simulation.}
\label{fig:herd}
\end{figure}

The first sexual reproduction event is usually between genotypically and phenotypically similar (if not identical) agents.  As a result most of the agents created in the early stages have a high genetic similarity (depending on random mutations).  Since they also all tend to be born in the same part of the environment they also tend to select the same or similar paths from their genome to express.  Thus, the pair of identical movers quickly becomes a herd.  Growth of the herd is usually exponential until the resource limits are reached and it cannot be sustained without mutation leading to emigration (see below).

Figure~\ref{fig:herd} shows the herd structure in one simulation as a heat map of phenotypic differences.  Each row (and column) represents an agent.  Each intersection in the heat map represents the difference between the daily activities of the two agents.  The axes have been ordered the same way so the diagonal shows comparisons between an agent and itself.  The ordering of agents has been selected by performing single link hierarchical clustering on the matrix of differences.  Such a clustering groups similar agents together.  In the heat map, black represents identical daily activities of the two agents and white expresses the maximum difference between two agents in the population. Shades of red show differences in between these extremes.  Since similar agents are grouped together each large dark square around the diagonal represents a group of agents with the same or similar daily activities.  These groups are interpreted as herds.

\begin{figure}[!t]
\begin{center}
\includegraphics[width = 240pt]{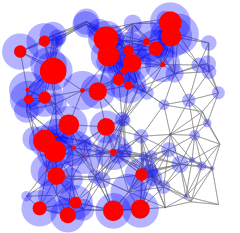}
\end{center}
\caption{An example network of sites indicating the number of visits (blue) and occurrences of sexual reproduction (red) per site.  The radius of the circle indicates the frequency on a logarithmic scale.}
\label{fig:sites}
\end{figure}

These herds are also observed to be philopatric from a number of different perspectives.  It is common for agents to focus on one area of their genome over their life often repeating the same locally optimal path forwards and backwards.  This can be seen as a strong territorial fidelity but is also an artifact of the simulation design.  Since agents in the same herd will all repeat the same path this can also be seen as the territory of the herd or sub-population.  This territorial social coordination is carried over in the herd from one generation to the next.

Breeding site fidelity within agents and within territorial herds is observed.  Which sites are used by the herd emerge as social contingencies, but after being established become characteristics of the herd's social structure.  Figure~\ref{fig:sites} shows the example network from Figure~\ref{fig:map} displaying the breeding sites used in a 10000 day run on that network.  A dozen sites have been used as the primary breeding sites with other sites being used less frequently and many sites not being used for breeding at all.  The frequency is on a logarithmic scale.

\begin{figure}[!t]
\begin{center}
\includegraphics[width = 240pt]{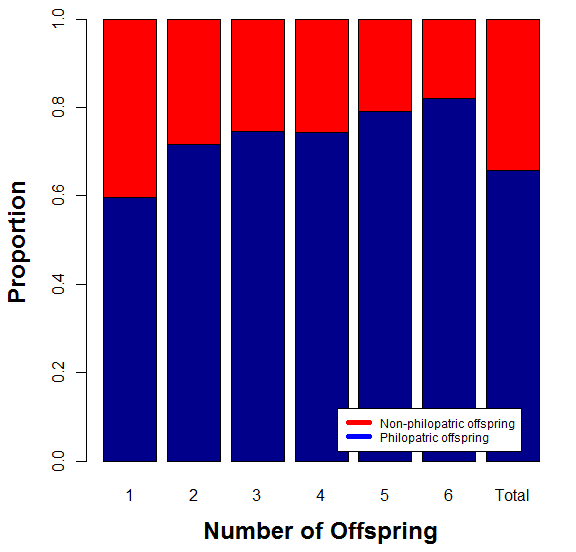}
\end{center}
\caption{Percentages of philopatry among breeders in the population by number of children.}
\label{fig:phil}
\end{figure}

Among the common breeding sites are the birth sites of many of the herd members.  An idealized herd might have a single breeding site at which all members of the herd were born.  This herd would display a pure form of natal philopatry.  Although an idealized herd is not observed there is evidence of natal philopatry in many agents.  Figure~\ref{fig:phil} shows that 66\% of parents engage in natal philopatry for some of their breeding events.  Natal philopatry is also more common in agents that have more children. 

% breeding site map and log plot

% philopatry chart(s)

Herding and philopatry are supported by a high degree of phenotypic and often genotypic similarity between agents in the same herd or using the same breeding site \cite{MC15}.  This leads to a tendency for agents within herds to breed with other agents from the same herd.  This implies a large amount of phenotypic similarity should be expected between parents.

Levenshtein edit distance was used to measure the similarity between two paths in the network (representing the daily activities of the agents).  Agents that have the same daily activities will have a measurement of zero (most similar).  Each addition, deletion, or substitution of a site will increase the measurement of distance.  Parents must have met sometime during the day to breed.  Thus all parents must have at least some degree of phenotypic similarity.

\begin{figure}[!t]
\begin{center}
\includegraphics[width = 240pt]{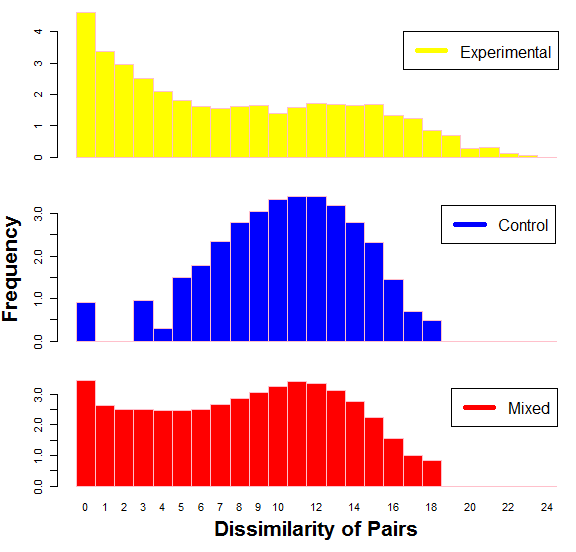}
\end{center}
\caption{Histogram of parents by difference of daily activities.  Top: All parents in runs of length 10000 days.  Middle: Control agents breeding with each other. Bottom: Control agents breeding with their offspring.}
\label{fig:phe}
\end{figure}

Figure~\ref{fig:phe} is a histogram of the similarities of daily activities between parents.  The top histogram shows the parents from over 100 runs of length 10000 on a logarithmic scale.  The left most rank indicates parents with no difference between their daily activities.  It can seen that the number of parents with no difference between them is orders of magnitude larger than even those with a single site difference in their daily activities.  This frequency continues to decline as difference increases.  The frequency levels out before dropping off completely near the maximal difference.  Parents with large differences between them likely represent members of different herds that meet by chance and this data indicates that these meetings still occur with some regularity.

The shape of this distribution contrasts with a control group (see below) where the network is seeded with 100 random agents.  Like the seed agents described above, their genomes (and thus daily activities) are generated by a random walker.  This represents a case where agents have no genetic or phenotypic similarity to rely on to find mates.  The middle histogram of Figure~\ref{fig:phe} shows the difference between parents that were lucky enough to find one another in this control group.  The control agents when breeding with one another display a near normal distribution around the mean difference in the population.

After these control agents found one another they created offspring that were more genetically similar to their parents.  The initial control agents would frequently breed with these offspring (ostensibly taking advantage of the genetic similarities).  The bottom histogram of Figure~\ref{fig:phe} shows the differences among parents that had one initial control agent and a second agent that was a descendant of the initial control agents.  Note that the shape of this histogram appears to be a hybrid of the shapes of the other two classes.

% log plot of phenotypic similarity

The above analysis is repeated for genetic difference between parents.  In simulation genomes are typically much longer than daily activities.  A typical seed agent has a genome that is four times longer than a typical phenotype and after evolution population genomes usually range from the length of a typical phenotype to one thousand times the length of a typical phenotype in extreme cases.  As a result we were curious if genetic assortment was also occurring due to the phenotypic assortment.

% log plot of genetic similarity
\begin{figure}[!t]
\begin{center}
\includegraphics[width = 240pt]{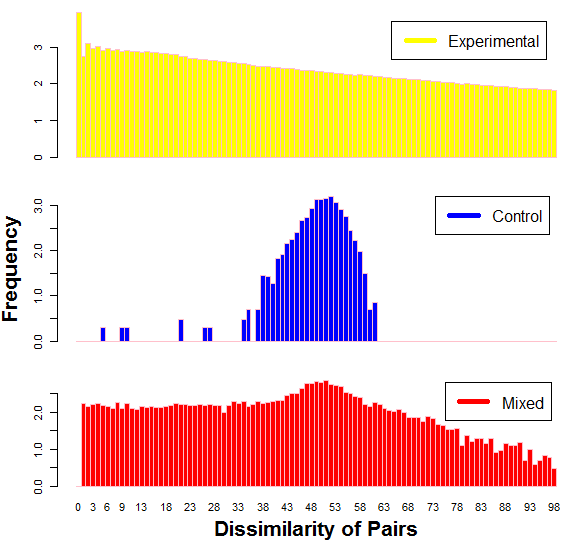}
\end{center}
\caption{Histogram of parents by genetic difference.  Top: All parents in runs of length 10000 days.  Middle: Control agents breeding with each other. Bottom: Control agents breeding with their offspring.}
\label{fig:gen}
\end{figure}

Figure~\ref{fig:gen} shows that there is a genetic assortment among the agents as well.  The top histogram represents the genetic similarity between parents in 10000 day runs.  This histogram shows a clear exponential decay in parent frequency as genetic difference increases (recall these plots are on a logarithmic scale).  There is a gentler decay than when comparing phenotypes, but there is no leveling off.  Part of this is due to genetic drift observed in the herds as time goes on \cite{MC15}.

As with the phenotypes, the contrast to the control population of random agents (middle histogram in Figure~\ref{fig:gen}) shows a near normal distribution of parents around the mean of genetic difference among the random control population.  Also similar to the phenotypic analysis, when the control population breeds with their descendants they begin a trend of assortment that is similar to the population of non-random agents (bottom histogram in Figure~\ref{fig:gen}).

We propose that the primary mimetic process at work in these observations is common descent.  Initially, the agents are all genetically similar because they are all copies of the first agent except for a few possible mutations.  The coordination of behavior of the first few agents is ensured by their similar genetic heritage and that they occupy the same environment.  As the simulation progresses and variation increases in the population common descent still plays a major role in coordinating the behavior of herd members.

Philopatry is also largely maintained through common descent.  Breeding and birth sites are maintained in populations over many generations because the members that use those sites have maintained the genetic similarity that led to that site being used in the first place.  Maintenance of this site as a breeding location is much easier than evolving as a group to use a new site.

The assortative mating patterns emerge for the same reasons.  Initially all herd members are genetically alike so assortment is guaranteed.  This is upset as variation is introduced by mutation and recombination.  When a novel population emerges it either has migrated to a new spatial region or it has become out of sync with the breeding strategies of the old herd.  In both cases it is more likely for herd members to breed with members of their own herd than others just because of proximity.  The phenotypic assortment is assured and the genotypic assortment either exists naturally or is created after multiple generations of phenotypic assortment.

\subsection{Control Population}

The typical run in the simulation begins with a single seed agent that is randomly generated.  In contrast to these runs, a control population is run with one hundred initial random agents.  Each agent in this control population is randomly generated and thus all similarity is due to random chance.  No agents in this initial population share common descent.  It is presumed that all mating that occurs between these agents occurs due to a convergence of their breeding behaviors.  Since these agents have not evolved, their common behaviors are due to a random convergence.

In the experimental runs described above the first sexual reproductive events come about through common descent.  In control runs common descent can play no role.  Instead of two agents traveling together, agents have chosen similar breeding sites (and times) coincidentally.  A large number of agents in the control population have no offspring at all because they could not find partners.

The control population has the potential to breed either among itself or, like in the experimental runs, they can breed with their offspring.  Offspring will regularly be more phenotypically similar to their parents than others from the initial population due to common descent (as above).  Comparing the mating behavior of the initial population with other members of the initial population provides a distribution around the mean difference between agents in that population.  This is the standard hypothesis of a random mating strategy.  This strategy is quickly erased with their offspring in favor of an assortative strategy even among members of the initially random population.

After many agents in the control population have died out, it appears very similar to what is seen in the single agent seed runs (though usually at later stages).  The sites where the coincidental meetings occur become likely locations for future breeding because of common descent.  After the seed population is gone, all that can be seen are herds forming around the locations of the coincidental meetings.  While initially varied, the members of the herds grow genetically more similar over generations due to assortment, which leads to a very similar dynamic as observed in experimental runs.

\subsection{Mutation and Emigration}

Another main difference between the experimental and control runs is the role that mutation plays in migration.  Initializing the simulation with multiple agents will ensure that these agents cover more resource sites than a single agent.  This gives more chances for success and so the population that emerges from the control population also tends to cover more resource sites.  Note not all sites are guaranteed to be used as coincidental meetings dictate which and how many sites are eventually used by the herds.

In an experimental run, agents migrate to a different set of sites through mutation.  For this to happen two conditions must be met.  First, there must exist a sub-path of an agent's genome that maximizes resource gain better than the one currently used by the agent.  Second, there must be a means for the agent to exploit this better region by moving through the environment to the new sites.  

The first condition is commonly satisfied.  A mutation in an unused portion of the genome could optimize that region increasing its resource gain.  A mutation could occur in the used portion of the genome causing it to decrease its resource gain.  The portion of the genome that is used by the agent is only guaranteed to be locally optimal and there is no reason to assume it is globally optimal.  

The second condition is more challenging to achieve and we consider two cases.  The easier case is when the new region of the genome and the old region of the genome share some overlapping sites.  These sites can then serve as a way for the agent to move in its genome without having to move in the environment.  When this type of adaptation occurs the new behavior will commonly still overlap with the old behavior and so members of the new herd and the old herd may interbreed.  The more difficult case is when the new region of the genome and the old region of the genome do not share any overlapping sites.  In this case there must also exist a region of the genome that can serve to bridge the two regions.  We know this region exists because the genome is a single path, but in many cases the bridge may be smaller than the actual genetic path that links the two regions (because of overlapping sites as above).  However, for the bridge to be used it must also be a region that is of greater resource gain to encourage the agent to leave its local optimum.  This will often require a large number of fortuitous mutations and is not often observed in the simulations.  However, due to the large number of agents found over the course of the simulation, populations do migrate in this fashion.

Migration of agents that results in successful populations often depends on the topology of the underlying random geometric network of sites.  If two areas are separated by too few paths a genome may never evolve to cross into the isolated area.  This is because mutation of the path in the network is modeled as a random walker on the network and random walkers will tend to accumulate in areas of higher edge density \cite{L93}.

\section{Discussion}

We attribute the successful sexual reproduction in the agents to non-social mimetic processes of convergence and common descent.  Specifically common descent seems to be a strong force in the simulation ensuring the fidelity of behavior within breeding populations.  Convergence seems to be a force that leads to the meeting and genetic exchange between populations that is important in maintaining genetic variation.

We believe that these non-social evolutionary forces will lead to similar mating dynamics whenever two conditions are satisfied.  First, the occurrence of mating between members of a population depends on the spatial arrangement of members of the population.  This is the social coordination problem (called the encounter problem by Gimelfarb \cite{G88}).  This requirement seems very ubiquitous in the natural world, but is often omitted in simulation.  Second, the genotype must influence the spatial arrangement of members of the population.  If it does not, then agents are merely random walkers and this random behavior cannot be selected for or against.  This influence can be direct or indirect requiring environmental interaction.

In the agents, the genome has a strong influence on the spatial distribution as it encodes paths in space directly.  We expect that other types of genomes may also lead to these social organizations.  In particular, we have considered another type of agent that encodes transition probabilities to move to a new site based on its current site.  We would expect that these agents might evolve to display herding, philopatry, and assortative mating and we plan to test this important hypothesis in the future as we also encourage others to test this hypothesis.

Finding evidence that supports this hypothesis would help to explain the ubiquity of sexual reproduction in the natural world even in the absence of social mechanisms of coordination of individuals.  All members of sexual species (animal or non-animal) must coordinate their sexual reproduction in real space.  So we must assume that our first condition holds for these species.  It is reasonable to assume that the genomes of many forms of life are attuned to certain environments.  Even if the population in question is not capable of locomotion the survival rates of individuals exposed to different environments could lead to differential distributions of individuals by their genotypes and their phenotypes.  We suggest that this is enough to support the types of organization seen in our agents.  However, we also believe that these behaviors exist on a spectrum and that populations that have less genetic influence on spatial distribution will likely display these behaviors also to a lesser degree.  All that is necessary for a population to survive is that these behaviors are strong enough to support continued reproduction.

We also believe that non-social mimetic processes can lead to social mimetic processes because our simulation shows that social organization can emerge from non-social processes in sexually reproductive populations.  In particular, the organizational abilities of herding, philopatry, and assortative mating all ensure that a certain social structure emerges in the population.  This social structure can then serve as the environment in which social interaction can evolve.  This might mean that the non-social mimetic processes that lead to social organization may be replaced by social mimetic processes that support that same or similar organization.

For instance, a herd that has emerged due to non-social mimetic processes provides the environment in which agents are regularly exposed to other members of the herd.  We have seen that mutation may take an agent away from the herd in these cases.  While this may lead to successful emigration, it often leads to death.  Agents that become socially inclined to stay near the herd would be insulated from the negative effects of these mutations and might not be selected against in some environments.  As a result the herd can move from being non-socially organized to being socially organized (likely shifting the dynamics of the population).

This is one possible case of what we call a scaffolding of social behaviors around sexual reproduction (Dennett\cite{D95} calls this phenomena ratcheting).  The scaffolding process would provide a common pathway for populations to evolve social behaviors from non-social ones due to the constraints of sexual reproduction.  If such a scaffolding effect is possible we would expect it to be present in a wide range of sexual organisms.  It may be the process that has led to the many examples of social processes in nature including human beings.  As future work we intend to expand the simulation discussed to test this hypothesis by exploring conditions under which a scaffolding of social behaviors around sexual reproduction can be encouraged or discouraged.  

\section{Conclusion}

We have argued that under two conditions a non-social evolving population has the means to organize itself for the purposes of solving the social coordination problem of sexual reproduction. Given that sexual reproduction in a population is constrained by the spatial distribution of individuals, and as long there is a genetic influence on the spatial distribution of individuals, we suggest social coordination will emerge around sexual reproduction.

This social organization's only purpose may be to support sexual reproduction and may remain non-social in origin in many populations.  However, we suggest that in some populations this social organization can serve as scaffolding of social behaviors by providing the social environment for these behaviors.  

%\end{document}  % This is where a 'short' article might terminate

%ACKNOWLEDGMENTS are optional

\bibliographystyle{abbrv}
\bibliography{main}  % sigproc.bib is the name of the Bibliography in this case
% ACM needs 'a single self-contained file'!
%\balancecolumns

\balancecolumns
% That's all folks!
\end{document}